# Neurophotonics beyond the Surface: Unmasking the Brain's Complexity Exploiting Optical Scattering

Fei Xia[1,*], Caio Vaz Rimoli[1,2,*], Walther Akemann[2], Cathie Ventalon[2], Laurent Bourdieu[2], Sylvain Gigan[1], Hilton B de Aguiar[1]

[1] Laboratoire Kastler Brossel, ENS-Université PSL, CNRS, Sorbonne Université, Collège de France, 24 rue Lhomond, 75005 Paris, France

[2] Institut de Biologie de l'ENS (IBENS), École Normale Supérieure, CNRS, INSERM, Université PSL, Paris, France

* These authors contributed equally.

Corresponding authors: F.X. fei.xia@lkb.ens.fr; C.V.R. caio.vaz-rimoli@lkb.ens.fr; W.A. akemann@biologie.ens.fr; C.V. cathie.ventalon@bio.ens.psl.eu; L.B. lbourdieu@biologie.ens.fr; S.G. sylvain.gigan@lkb.ens.fr; H.B.A. h.aguiar@lkb.ens.fr

## Abstract

The intricate nature of the brain necessitates the application of advanced probing techniques to comprehensively study and understand its working mechanisms. Neurophotonics offers minimally invasive methods to probe the brain using optics at cellular and even molecular levels. However, multiple challenges persist, especially concerning imaging depth, field of view, speed, and biocompatibility. A major hindrance to solving these challenges in optics is the scattering nature of the brain. This perspective highlights the potential of complex media optics, a specialized area of study focused on light propagation in materials with intricate heterogeneous optical properties, in advancing and improving neuronal readouts for structural imaging and optical recordings of neuronal activity. Key strategies include wavefront shaping techniques and computational imaging and sensing techniques that exploit scattering properties for enhanced performance. We discuss the potential merger of the two fields as well as potential challenges and perspectives toward longer term *in vivo* applications.

## Introduction

The brain acts as the central regulator in all vertebrate and most invertebrate organisms[1]. Comprehensive study of its structure and function is not only paramount to our scientific understanding but also crucial for developing interventions for brain-related pathologies[2,3]. In this context, the field of neurophotonics, a domain that capitalizes on optical tools to study the nervous system (Figure 1), has emerged as a powerful strategy for brain studies. Three defining strengths of optical approaches include: (i) their minimal invasiveness [4–6], (ii) their enhanced specificity when combined with molecular labeling[7–11] or label-free optical techniques[5,12–16], allowing for targeted imaging at cellular and molecular levels, and (iii) the possibility of chronically recording the same structures of interest, such as neurons, dendrites, and spines[17], during development, learning, and sensory deprivation[18,19]. However, there are persisting challenges that limit the comprehensive use of optical techniques in brain research. In this perspective paper, we specifically focused on optical imaging and sensing tools to probe the brains of animal models. Neuroscientists have proposed a key objective for optical probing of the brain: to develop and integrate advanced optical probing techniques that offer high spatiotemporal resolution, large-scale recording and mapping of neural activity while ensuring safety and minimal invasiveness[20–22]. Meeting this objective necessitates advancements in: 1) Probing depth, especially important given the size variations of the brain, from larger scales in humans to smaller scales in other species[7,23–25] (Figure 2a). 2) Expanding the field of view (FOV), allowing for a more holistic capture and understanding of neuronal networks[26–28] (Figure 2d). 3) Improving probing speed to capture and interpret dynamic biological activities in both 2D and 3D contexts[9,29–32] (Figure 2b). 4) Ensuring biocompatibility: minimizing phototoxicity and avoiding damage from implanted devices, thereby preserving the brain's structural and functional integrity during investigations of the brain using optical methods[33,34] (Figure 2c).

Here, we review recent advances in techniques and devices popularized in the complex media community that have begun to show promise in addressing some of the key challenges (Fig. 2) and discuss our perspectives on moving forward for *in vivo* applications.

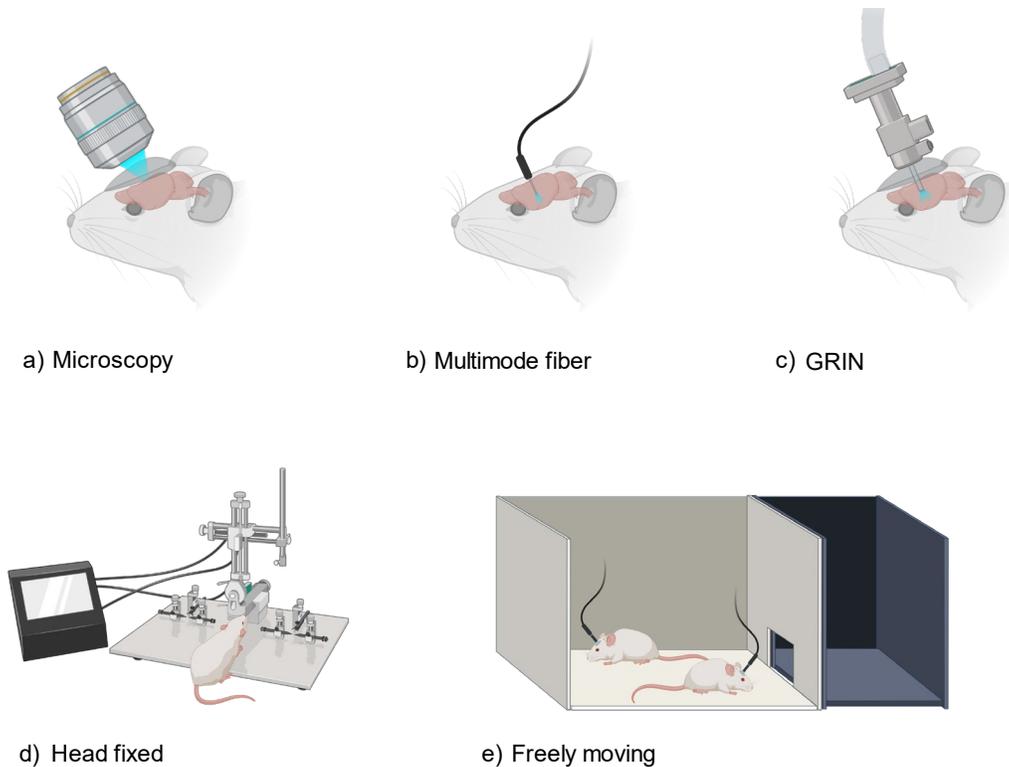

**Figure 1**. Overview of Diverse Brain Probing Techniques: **(a)** Microscopy: Traditional imaging with direct optical access to the brain. **(b)** Multimode Fiber: Flexible approach using a fiber optic cable for light delivery and signal collection. **(c)** GRIN (Gradient Refractive Index) Lens: Minimally invasive imaging through a small-diameter lens. **(d)** Head Fixed: Apparatus for stable imaging with restrained subject movement. **(e)** Freely Moving: Setup allowing for natural behavior during imaging with a mobile recording system. Figures (a-e) adapted with BioRender.com.

## Opportunities: Bridging the Gap

The complex media field studies light propagation in materials with highly inhomogeneous optical properties. Tools developed in this area include advanced computations on light scattering in optically heterogeneous micro- media and algorithm design for shaping light through diffusive materials and image recovery using scattering information[35]. While rooted in fundamental light scattering, its implications naturally extend to neurophotonics, due to the highly scattering nature of brain tissues.

Key techniques in the complex media field can be broadly categorized into two groups: wavefront shaping[35–38] through complex media and computational imaging and sensing techniques using complex media[39]. Wavefront shaping, a technique that modulates the phase and amplitude of incoming light waves using light shaping devices such as spatial light modulator (SLM), is emerging as a promising avenue. Adaptive Optics (AO), a wavefront shaping method focused on spatiotemporal resolution across various optical imaging modalities[40–54]. Looking ahead, recent wavefront shaping techniques that address scattering (higher-order light distortion)[46,55–64] have the potential to further improve signal and resolution, especially at depths where scattering becomes a critical limitation (Fig. 1a). Recent insights into local correlation during scattering events, i.e. the memory effect[65], in chromatic[66,67], shift[68], tilt/angular[69,70], and others[71] may guide more efficient light manipulation deep within tissues.

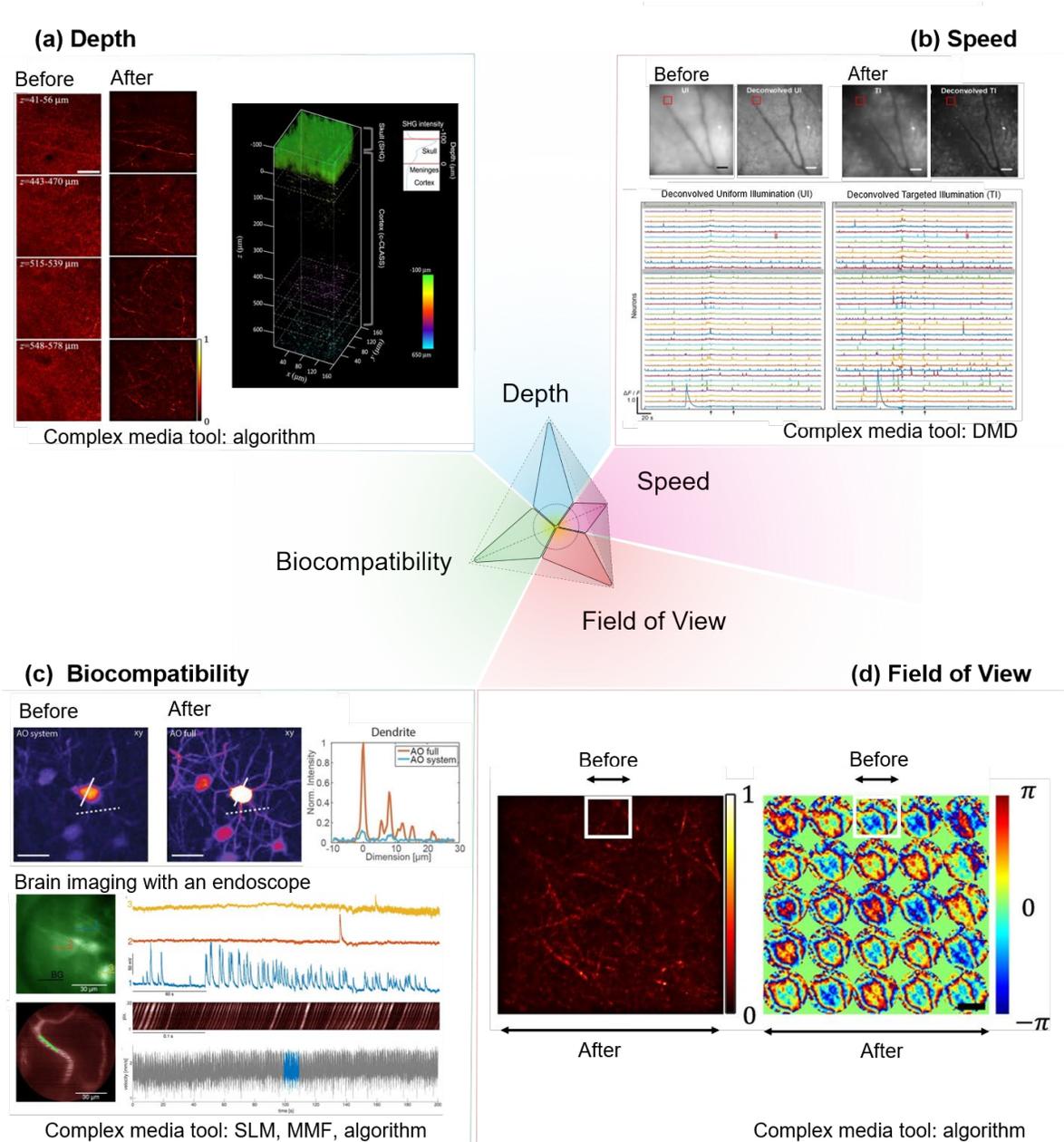

**Figure 2. Representative advances from tools commonly used in the complex media community to address challenges in optical probing of the brain**. **(a) Depth**: scattering and aberration compensation using computational techniques to enhance reflectance imaging of cortical myelin through the skull in the living mouse brain[114]. Before: conventional reflectance microscopy through the mouse skull. Left panel: After: computational conjugated adaptive optical corrected reflectance microscopy of cortical myelin in the mouse brain through skull. Right panel: 3D reconstruction of label-free structural information through skull. Scale bar: 40 µm. **(b) Speed**: Fast 3D volumetric imaging with targeted illumination of neurons in the mouse cortex labelled with a calcium indicator (GCaMP6f) to increase signal-to-noise of recorded neurons. Before: conventional volumetric calcium imaging with electrically tunable lens and extracted traces after deconvolution. After: illumination-targeted volumetric calcium imaging and extracted traces after deconvolution[30]. Scale bar 50 µm. **(c) Biocompatibility**: upper panel: enhanced signal given the same laser power enabled by adaptive optics[54]. Before: low signal-to-background of fluorescence-labelled neurons in the hippocampus around 1 mm depth imaged transcranially by conventional three-photon fluorescence microscopy. After: high signal-to-background neurons in the hippocampus imaged by adaptive optics. Scale bar: 20 µm. Lower panel: brain imaging of deep subcortical neurons labeled with a genetically-encoded calcium indicator GCaMP6s using a multimode fiber-based endoscope combined with wavefront shaping for minimally invasive imaging[115]. Scale bar: 30 µm. **(d) Field of view**: enlarged field of view with diffraction-limited high-resolution imaging enabled by computational conjugated adaptive optics (after) compared with computational adaptive optics without conjugation (before, white boxes)[114] Left: Image of myelin. Right: Phase pattern for aberration correction. SLM: spatial light modulator, DMD: Digital Micromirror Devices, MMF: multimode fibers. Panel (a) adapted from ref[114] under license CC-BY 4.0. Panel (b) adapted from the ref[30] under license CC-BY 4.0. Panel (c) the top images adapted from ref[54] and the bottom images adapted from ref[115] under license CC-BY 4.0. Panel (d) adapted from ref[114] under license CC-BY 4.0.

The memory effect refers to the phenomenon where the optical fields of scattered light remain correlated when certain properties of the light, such as position, wavevector direction, polarization, or spectrum, change over a specific range. As illustrated in Figure 3: 'Chromatic' refers to changes in the light's wavelength; 'shift' pertains to the displacement or angular deviation of light beams; and 'angular or 'tilt'' involves changes in the direction of light propagation. Memory effect enables the prediction of how light's properties change with scattering, facilitating computational or hardware-based tools for enhanced imaging quality through scattering tissues or interfaces with complex optical properties, such as multimode fibers. For instance, by conjugating the light modulation plane to specific locations within the scattering medium (Fig. 2d), we might find an optimal balance between enhancements of the signal intensity, FOV, and spatial resolution[72,73]. In the brain, despite the relative dense packing of neurons and vasculature, fluorescence microscopy often reveals a sparser distribution particularly when given at a certain color channel, a result of selective fluorescent labeling targeting specific cellular or vascular components or a sparse expression of fluorescence[25,74–76]. Leveraging sparse and compressive sampling or scanning techniques, like Acousto-Optic Deflectors[77–80] (AOD) and Digital Micromirror Devices[30,81–86] (DMD) (Fig. 2b), ensures efficient photon utilization. Such methods not only expedite the imaging process but also preserve the photon budget, setting the stage for up to one order of magnitude increase in the imaging speed (Fig. 2b) and reduction in the laser power (Fig. 2c) for faster and physiologically safer recording given proper guide stars for wavefront shaping[87]. Guide stars in imaging are akin to its astronomical counterpart; they serves as a reference light source from various contrast mechanisms, such as harmonic, photoacoustic, fluorescence, and scattered light[87], within the sample to facilitate the correction of light distortion caused by scattering. By employing the guide stars, we can guide the wavefront shaping process to more precisely manipulate incoming light waves. This improves the efficiency of the photon budget of the incoming field in enhancing the focus intensity and signal at greater penetration depths of imaging systems and in reducing laser intensity, minimizing potential photodamage to biological tissues. They could even help capture faster events such as millisecond action potentials in neurons[9,29,30,88–90] (Fig. 2b). Furthermore, compressive random-access sampling with fast light modulators, like AODs, permits to integrate fast temporal sampling and wavefront shaping[79,88,89], including adaptive correction of aberrations and scattering over an extended FOV that effectively exceeds the range of the angular memory effect, by taking advantage of the fast AODs' update rate to correct multiple local aberrations almost synchronously with the progression of a scanning beam, whether in pixel-by-pixel or random-access scan mode[91]. The primary advantage of employing wavefront shaping in enhancing the capabilities of state-of-the-art optical microscopy lies in the optimization of the photon budget. This technique enables the strategic redistribution of photons to either augment the imaging speed or expand the field of view (FOV), all while maintaining a fixed photon allowance for biological imaging under safe physiological conditions. However, challenges remain in terms of its shaping speed, which needs to be improved to overcome the temporal decorrelation of the scattered light field (Fig. 3h).

In the realm of computational imaging and sensing techniques through complex media, image reconstruction[39,92,93] and signal processing[94–96] methods that exploit random or scattering media properties have emerged as potential game-changers. Leveraging the inherently locally correlated nature of scattered light, techniques such as auto-correlation[97], cross-correlation[98,99], and patch-connecting-based[100] image reconstruction methods have been proposed. These aim to directly reconstruct images through highly scattering media, with the potential to achieve a larger FOV at greater depths. Image reconstruction fundamentally is a process of solving optimization problems, which can be categorized into convex and non-convex cases. Convex optimization problems are generally more straightforward to solve because their global minima are easily identifiable. On the other hand, non-convex optimization problems, more common in imaging through scattering tissue, often suffer from multiple local minima, complicating the search for the global minimum. In this challenging landscape, deep learning[101,102] emerges as a powerful tool, offering robust methods that learn from data to effectively approximate global optima — opening new exciting avenues for neurophotonics imaging, with significant potential to enhance its capabilities. It provides not only an alternative tool for solving optimization problems, such as optimization as unrolled neural networks[98,103,104], but also enhances image reconstruction with deep learning models for better generalization of the scattering problems[101]. On the other hand, the analysis and understanding of speckle—a highly sensitive interference pattern commonly seen when light propagates through complex media[105,106]—has proven to be extremely powerful and promising. In the brain, the detected speckle signal can be highly sensitive to various events, such as calcium signaling[107], an indirect indicator of voltage fluctuations, and blood flow[108].

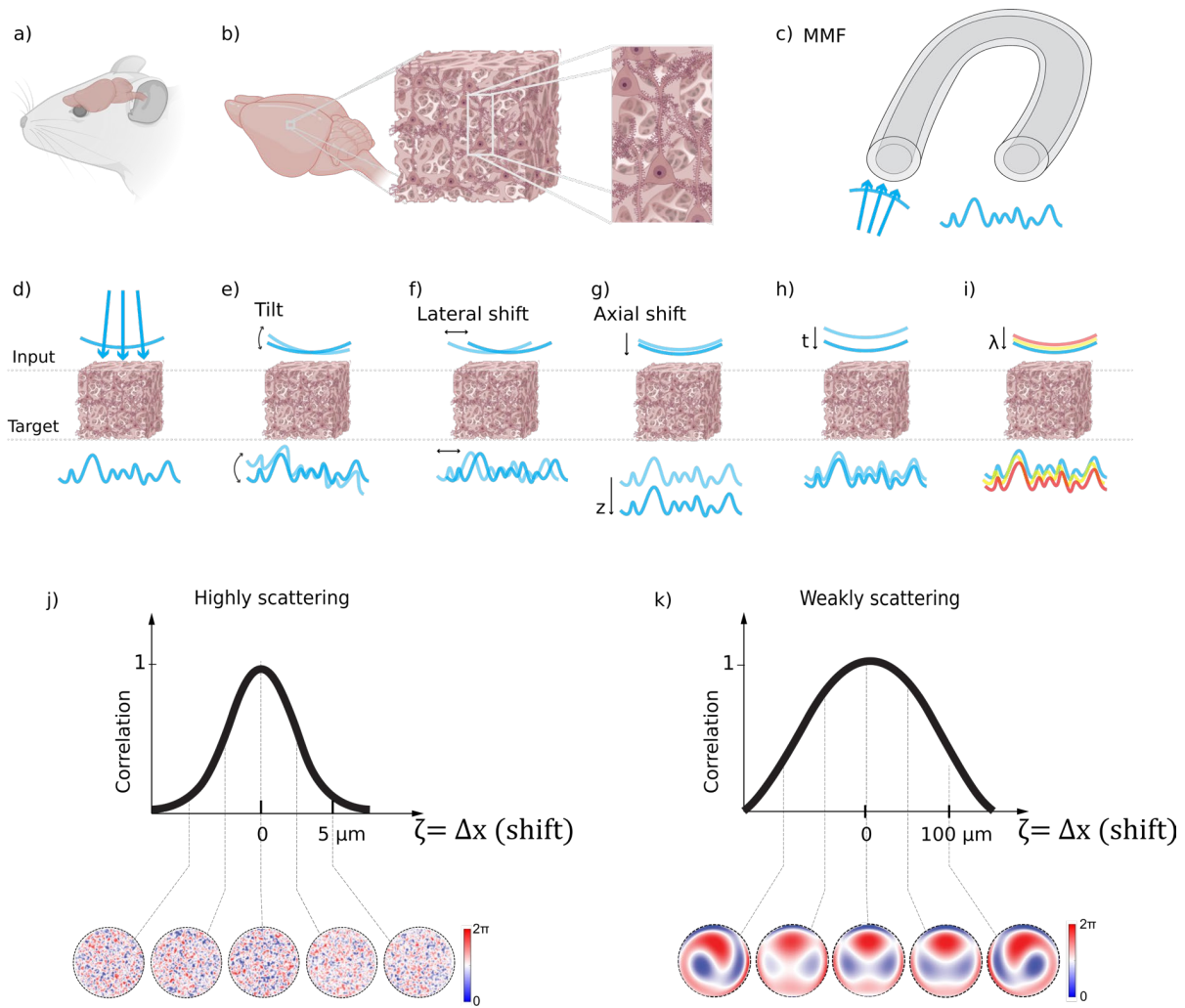

**Figure 3**. Optical access to the mouse brain through a scattering medium: **(a)** Schematic of a live mouse highlighting the brain area; **(b)** Inhomogeneous structures within the mouse brain that can cause optical scattering; **(c)** Multimode fiber (MMF), a frequently studied complex scattering medium in complex media field, is also often utilized for optical access to the brain; **(d)** Scattering-induced wavefront distortion; **(e-i)** Various memory effects: **(e)** Tilt/angular memory effect; **(f)** Lateral shift memory effect; **(g)** Axial shift memory effect; **(h)** Temporal memory effect; **(i)** Chromatic memory effect; **(j)** Representative quantitative correlation of wavefront correction pattern for achieving diffraction-limited focusing/imaging in highly scattering brain tissue, demonstrating that the range of the memory effect (defined by the full width at half maximum of the correlation curve) is substantially narrower compared to less scattering scenarios as shown in **(k)**. The patterns for correcting wavefront distortion in highly scattering media **(j)** are more complex than in weakly scattering media **(k)**. Note in (j,k), ζ could be any of the types of memory effect above in (e-i), but for the illustrative example we chose ζ = Δx (shift). Figures (a-b, d-i) adapted with BioRender.com.

Computational imaging techniques have shown promising enhanced results in brain imaging. For example, advanced signal processing methods such as non-negative matrix factorization (NMF), have proven instrumental in calcium imaging experiments by effectively removing noise and isolating signal components[109,110]. Additionally, the utility of computational imaging extends to blood flow estimation and reconstruction from speckle patterns observed in brain tissues[111].

Furthermore, computational tools such as constrained NMF[112] and DeepCAD[113] have greatly enhanced denoising techniques and the ability to retrieve signals from significantly high noiselevels[112,113,116–118]. These advancements are particularly valuable for imaging through highly scattering tissues, where traditional imaging methods are challenged to provide clear and reliable data. Although still at its early stage, it is anticipated that computational imaging will continue to enhance the clarity and utility of acquired images, enabling more detailed and accurate studies of neural structures and functions in challenging imaging conditions.

**Challenges and Limitations Towards Longer-Term *In Vivo* Applications**

Although the progress mentioned earlier has been exciting, the further adoption of these for longer-term *in vivo* biological studies still faces challenges.

*In vivo* applications involve imaging and sensing activities within the brain of a living and behaving animal, which raise a first challenge in term of recording artifacts linked to movements. The most common *in vivo* strategy is to fix the animal's head under a microscope, allowing for a good control of the sensory stimuli applied to the animal, as well as to accurately measure its behavior. Neurophotonics techniques developed for *in vitro* samples can be adapted for head-fixed animals provided that motion is taken into account. This encompasses micro- to milli-meter scale motions from heartbeats and respiration, to blood flow, and bulk motions induced by body movements and muscle contractions. These motions cause spatiotemporal noise dynamics in the tissue's scattering properties. Temporally, these dynamics are observable down to the millisecond range, and spatially, they can be seen down to the micron level. For example, regarding the bulk motion of the brain, in the case of 2-photon imaging experiments in the cortex with a cranial window, motion artifacts were observed to be around 2 to 4 μm in axial direction, which is much shorter than the 150 μm thickness of the optical window[79]. When implanting an optical fiber, one expects to encounter similar motion artifacts when exploring shallow regions of the brain. Interestingly, however, fewer motion artifacts are observed when a fiber is implanted in deeper brain regions. Indeed, this has been observed for 2-photon imaging with GRIN lenses[119,120]. From a technical standpoint, this poses concerns regarding the stability and speed of wavefront shaping techniques, as well as noise issues in computational imaging and sensing techniques. These factors underscore the need for adaptive imaging solutions that can recalibrate in real-time, ensuring consistent performance[121]. However, these hurdles, though significant, are not insurmountable. The way forward may involve a co-design philosophy, harmoniously melding wavefront shaping systems, algorithms, and imaging systems. For head-fixed animals, introducing an 'animal-in-the-loop' design could be revolutionary. This innovative approach would use real-time feedback from the animal's physiological and behavioral changes to continually adapt the imaging process, such as using online motion tracking to adapt in real-time the scanning scheme[122] or the heartbeat signal to gate the optical signal and remove heartbeat-related imaging noise[54].

A second challenge is improving depth penetration in brain tissues. In brain imaging, the depth achievable with current technologies varies significantly across different microscopy techniques and contrast mechanisms. For example, we have summarized the depth penetration capabilities of some of the most popular fluorescence microscopy techniques, including one-photon, two-photon, and three-photon excited fluorescence, as follows (in the context of *in vivo* adult mouse brain imaging):

| Fluorescence microscopy | Demonstrated depths so far with high spatial resolution (close to diffraction-limited) | Potential estimated depth limits with high spatial resolution (close to diffraction-limited) |
|---|---|---|
| Excitation: One-photon excited<br>Detection: Widefield | 0.1 ~ 0.2 mm[123]<br>(visible range of light) | 0.6 ~ 0.8 mm[124,125]<br>(*near-infrared II or short-wave infrared light) |
| Excitation: One-photon excited<br>Detection: Confocal | 0.3 ~ 0.4 mm[126]<br>(visible range of light) | 1.5 ~ 2 mm[124,125]<br>(*near-infrared II or short-wave infrared light) |
| Excitation: Two-photon excited<br>(temporal focusing)<br>Detection: Wiefield | 0.3 ~ 0.4 mm[127]<br>(near-infrared I light) | 0.6 ~ 0.8 mm[124,125,128]<br>(*near-infrared II or short-wave infrared light) |
| Excitation: Two-photon excited<br>Detection: single-element detector<br>(e.g. photomultiplier tube, PMT) | 0.6 ~ 0.8 mm[7,129]<br>(near-infrared I light) | 1.5 ~ 2 mm[125,128]<br>(*near-infrared II or short-wave infrared light) |
| Excitation: Three-photon excited<br>Detection: single-element detector<br>(e.g. photomultiplier tube, PMT) | 1.2 ~ 2.1 mm[129,130]<br>(near-infrared II or short-wave infrared light) | 3 ~ 4 mm[125,131]<br>(*near-infrared II or short-wave infrared light) |

Table 1: Current and potential limits of depth penetration capabilities of one-, two-, three-photon excited fluorescence microscopy. Visible range of light: 380 - 700 nm; near-infrared I light: 700 - 900 nm. *Indicates optimal imaging windows around 1300 nm and 1700 nm (in the region of near-Infrared II between 1000-1700 nm also called the short-wave infrared range in similar or even broader ranges in some definitions.

Conventional one-photon (1P) microscopy is limited to depths of approximately 0.3-0.4 mm due to light scattering and absorption in the commonly used visible range of light[124,126]. Conventional two-photon (2P) microscopy extends this depth to about 0.6-0.8 mm[7,128]. Conventional three-photon (3P) microscopy further increases imaging depth to 1.2-2.1 mm[129,132]. The potential depth limits (table 1, column 3) for these imaging methods can be estimated based on effective attenuation lengths[125] depending on the excitation and detection method[124,128,131].

Techniques such as wavefront shaping and computational imaging have been developed to mitigate scattering and aberrations, potentially enhancing imaging depth and resolution. These advancements enable more efficient light delivery and collection deep within tissues. Particularly, wavefront shaping can be and has been coupled with 1P-, 2P-, or 3P-excited fluorescence contrasts, thereby having the capability to extend the depth for each modality. For example, in a proof-of-concept multiphoton wavefront shaping experiment, an enhancement of at least one order of magnitude for the 2P signal and a two orders of magnitude gain for the 3P signal were observed[52].

One fundamental barrier is the depth beyond which even sophisticated light manipulation or computational imaging strategies become potentially impractical (refer to table 1, column 3) for diffraction-limited focusing and reconstruction. In such cases, minimally invasive fiber optics are the only viable option for high-resolution imaging, especially in scenarios requiring high mobility or minimal interference with the subject (animals). Devices such as miniature multimode fibers can be used to bypass the scattering of tissues. Endoscopes incorporate multimode fibers (MMF)[133–136] (Fig. 1b, 2c, 3c), as a relay between the animal and a benchtop wavefront shaping microscope, ensuring minimal invasiveness. Pioneering works, such as MMF-based imaging for mouse brains[115,135] and *in vivo* histology[137] as well as deep learning for image reconstruction through MMF[138,139], provide glimpses into the potential future of neurophotonics for deep brain imaging.

Advanced wavefront shaping techniques can also help to address a third challenge: imaging in freely behaving configuration, which provides access to a wider range of behaviors, such as social interactions and sleep. Freely-behaving imaging was achieved thanks to the use of wavefront shaping assisted endoscopes based on fiber bundles[140,141]. Another approach is miniatures microscopes[142–146] that allow measuring neuronal activity using conventional widefield[142,143], 2-photon[144,145] and 3-photon[147–149] imaging methods. However, combining these microscopes with wavefront shaping techniques will necessitate miniaturization of beam shaping devices. The ultimate goal would be the development of wireless miniscopes[150,151], freeing subjects from physical restraints and promoting natural behaviors.

On the other hand, computational techniques such as machine learning can generally facilitate a more robust search for the global optimum in non-convex optimization problems that exist in probing through scattering tissues. Major application directions involve 1) the reconstruction of high-fidelity images from scattered light patterns, effectively 'learning' the tissue's scattering properties to inversely map the captured signals back to their original, unscattered state; 2) denoising images during high-speed imaging or challenging imaging scenarios; 3) decoding the scattered field/patterns for biomedical insights; 4) predicting correction masks in wavefront shaping.

For instance, neural networks have been utilized to predict the unscattered light path, allowing for real-time correction of distorted images caused by tissue scattering[152]. This method has the potential to enhance the depth penetration and resolution of imaging modalities such as two-photon and three-photon microscopy, making it possible to visualize neuronal activity deeper within the brain with unprecedented clarity. At the same time, deep learning has enabled enhanced image quality in challenging imaging scenarios by image denoising, such as high-speed voltage imaging[153] and high-quality calcium imaging[113,117]. This advancement has led to improved neural activity traces, facilitating more accurate spike inference. Furthermore, deep learning models have been applied to interpret the speckle patterns resulting from coherent light scattering, extracting meaningful biological signals such as cerebral blood flow from noise and thereby facilitating non-invasive imaging techniques that can monitor brain dynamics[154]. Deep learning has recently also been applied in predicting scattering or aberration correction patterns in brain imaging[155].

Looking ahead, machine learning, with its strengths in generalization and robustness, can be invaluable. Algorithms supported by machine learning can process and interpret vast amounts of data rapidly, ensuring that researchers keep pace with the time-varying optical properties in *in vivo* environment. As neurophotonics delves deeper into uncharted territories, a symbiotic relationship between industry and academia becomes essential. Industrial stakeholders can develop faster, more stable shapers and sensitive detectors, while academia can push boundaries in algorithmic and system design innovations.

## Conclusion

In conclusion, the merging of complex media research with neurophotonics marks the beginning of an era brimming with significant potential and opportunities. Moving forward, there is a need for collaboration and innovation across different disciplines, such as algorithm development, machine learning, optics, and neuroscience. This interdisciplinary approach is essential for overcoming existing technical challenges and unravelling better mechanistic understanding of the brain in the unexplored regime. Given the recent advancements in computational imaging and sensing through complex media, wavefront shaping technology, machine learning tools, and the myriad of chemical and biological tools developed in neuroscience, we believe there lies a tremendous opportunity to synergize these diverse fields.

## Acknowledgements

The authors thank Ruth Sims and Yusaku Hontani for their feedback and proofreading of the manuscript. F. X. and S.G. acknowledge funding from Chan Zuckerberg Initiative (2020-225346); C.R.V and S.G. acknowledge funding from: HFSP project N°RGP0003/2020; W. A., C.V. and L.B acknowledge funding from Institut de Convergence Qlife (ANR-17-CONV-0005); NIH BRAIN Initiative (1U01NS103464); ANR (ALPINS ANR-15-CE19-0011, EXPECT (17-CE37-0022-01) - The program Investissements d'Avenir  ANR-10-LABX-54 (Memolife), ANR-11-IDEX-0001-02 (Université PSL) and ANR-10-INSB-04-01 (France-BioImaging infrastructure); C.V. acknowledge ANR MULTIMOD (OP19-121-OTP-01). S.G. acknowledges funding from NIH 1RF1NS113251 and senior fellowship from Institut Universitaire de France. S.G. and H.B.A. acknowledge funding from H2020 Future and Emerging Technologies (grant no. 863203). H.B.A. acknowledges funding from ANR COCOhRICO (ANR-21-CE42-0013).

## Code, Data and Materials Availability

Data sharing is not applicable to this article, as no new data was created or analyzed.

## Disclosures

The authors declare no conflicts of interest.